\documentclass[10pt,letterpaper]{article}
\usepackage[top=0.85in,left=2.7in,footskip=0.75in]{geometry}

\usepackage{amsmath,amssymb}
\usepackage{changepage}
\usepackage{textcomp,marvosym}
\usepackage{cite}
\usepackage{nameref,hyperref}
\usepackage[right]{lineno}
\usepackage[nopatch=eqnum]{microtype}
\DisableLigatures[f]{encoding = *, family = * }

\usepackage[table]{xcolor}

\usepackage{array}

\newcolumntype{+}{!{\vrule width 2pt}}

\newlength\savedwidth

\raggedright
\usepackage[aboveskip=1pt,labelfont=bf,labelsep=period,justification=raggedright,singlelinecheck=off]{caption}

\makeatletter
\renewcommand{\@biblabel}[1]{\quad#1.}
\makeatother

\usepackage{lastpage,fancyhdr,graphicx}
\usepackage{epstopdf}
\fancyheadoffset[L]{2.25in}
\fancyfootoffset[L]{2.25in}
\begin{document}
\vspace*{0.2in}

\begin{flushleft}
{\Large
\textbf\newline{Story Arcs, Success and Diffusion of Turkish TV Dramas} 
}
\newline
\\
Guanghui Pan\textsuperscript{1},
Selcan Mutgan\textsuperscript{2},
Abdul Basit Adeel\textsuperscript{3},
Abbas K. Rizi\textsuperscript{4,5,6}
\\
\bigskip
\textbf{1} Department of Sociology, University of Oxford, United Kingdom
\\
\textbf{2} Institute for Analytical Sociology, Linköping University, Norrköping, Sweden\\
\textbf{3} The Pennsylvania State University, State College, PA, USA
\\
\textbf{4} DTU Compute, Technical University of Denmark, Kongens Lyngby 2800, Denmark\\
\textbf{5} Copenhagen Center for Social Data Science, University of Copenhagen, Denmark\\
\textbf{6} Department of Computer Science, School of Science, Aalto University, FI-00076, Finland

\bigskip

* akari@dtu.dk
\end{flushleft}
\section*{Abstract}
Turkish TV dramas, renowned for their extensive episodes and captivating romantic narratives, have gained global popularity, leading to the central question of this study: what drives the virality of these series based on their scripts and emotional story arcs? This study explores the global popularity of Turkish TV dramas, examining their narrative structures and emotional arcs to understand their widespread appeal. It delves into the cultural and emotional factors that resonate with international audiences, highlighting Türkiye's significant role in the global entertainment landscape. Through a combination of sentiment analysis and narrative examination, the research offers insights into how these dramas create emotional engagement and cultural connectivity, contributing to Türkiye's growing influence in global media.

\section*{Author summary}
We discuss the significance of narrative structures in the global diffusion of Turkish TV dramas. By analyzing emotional arcs and mapping them to audience reception data, we demonstrate that emotionally charged narratives, particularly those with a Rise-Fall arc, are highly successful in engaging international audiences. Cultural proximity plays a role in the initial diffusion, but the universal appeal of emotional storytelling helps transcend these boundaries. This study contributes to the growing field of narrative analysis within cultural diffusion frameworks, emphasizing the role of emotions in the success of time-based cultural products like TV dramas. Future work could explore more granular narrative features and their influence on audience engagement and cultural transmission.

\section*{Introduction}
The burgeoning success of Turkish television series, colloquially known as ``Türk dizileri," marks a pivotal evolution in modern popular culture, underscoring Türkiye's expanding sway within the international entertainment sphere \cite{berg2023turkish}. The narrative frameworks employed in storytelling mediums like films or TV series can be dissected through the lens of their emotional trajectories, mapping the variations in emotional intensity over time. This analytical approach posits that every narrative possesses a distinct ``shape,'' delineated through a time series depicting its fluctuating emotional valence \cite{vonnegut1999palm}. Such a methodology quantifies the ``happiness'' levels within a narrative, offering a lens to decode its emotional fabric. By employing this technique across a spectrum of narratives from specific cultural or genre-based cohorts, it's possible to unearth prevalent emotional motifs, thereby classifying narratives into distinct emotional archetypes. This method illuminates the pivotal trends in storytelling that forge emotional connections with audiences, transcending conventional cinematic elements. The seminal work by Reagan et al. \cite{reagan2016emotional} further enriches this discourse through an in-depth exploration of sentiment trajectories in narrative constructs.

The ascendancy of Turkish TV dramas on the global stage, especially from the onset of the 21st century, has been profound. Celebrated for their authentic depictions of Turkish ethos, these dramas have propelled Türkiye to the forefront as a cultural emissary. By the mid-2010s, Türkiye had ascended to become the world's penultimate exporter of television content, trailing only behind the United States. The allure of Turkish series is partly attributed to their episodic length and high production values, with each installment spanning between 120 to 150 minutes, mirroring the length of feature films and providing a deeply immersive narrative experience. The international acclaim of these series has not only spurred tourism in Türkiye, drawing fans to the locales of their beloved dramas but has also significantly bolstered Türkiye's cultural cachet on a global scale.

Turkish dramas boast a wide-reaching appeal, captivating audiences in diverse regions from the Arab world to Latin America and extending to Israel and Japan. This universal appeal is rooted in the socio-cultural themes that resonate with viewers across these regions. Intriguingly, the popularity of Turkish series in Israel persists despite geopolitical strains, suggesting that the cultural influence of Turkish television transcends political confines. The transformative impact of Turkish TV series has redefined Türkiye's stature within the global media landscape, positioning it as a formidable force. These series have emerged as potent instruments of soft power, shaping global perceptions of Türkiye, its populace, and its cultural heritage. This cultural phenomenon has garnered academic interest, with scholars probing the diverse factors underpinning the global allure of Turkish dramas.

Recently, Del Vecchio et al. delved into an analysis of 6,174 film scripts to decipher the correlation between emotional arcs and cinematic success \cite{del2021improving}. The study segmented these narratives based on their emotional trajectories, employing an econometric framework to forecast various success indicators, including box office earnings, IMDb ratings, accolades, and review volumes. Their findings highlighted the ``Man in a Hole'' arc, characterized by an initial downturn followed by an upsurge, as a harbinger of heightened financial returns. This discovery aligns with the findings of Reagan et al. \cite{reagan2016emotional}, which delineated six core emotional arc patterns within English literature. The insights from Del Vecchio et al.'s research are instrumental for content creators, offering a strategic lens to enhance narrative effectiveness and industry productivity \cite{del2021improving}. 

The research posits that the medium's duration (films versus literature) influences audience predilections for specific emotional journeys. For example, the Icarus arc  (rise then fall), marked by a pronounced emotional descent, finds greater resonance in literary formats, possibly due to the elongated engagement period, allowing readers to assimilate the emotional decline more fully. Contrastingly, only the ``Man in a Hole" narrative demonstrates a statistically significant association with elevated box office revenues, implying that films adhering to this emotional contour tend to outperform financially. Conversely, the Oedipus arc (fall, rise, then fall again), despite its prevalence, does not exhibit a marked impact on revenue generation \cite{del2021improving}.

Given the extensive episode lengths and enthralling romantic plots of Turkish TV dramas, this study seeks to unravel the underpinnings of their international virality, focusing on their narrative structures and emotional arcs. This investigation probes the widespread appeal of Turkish TV dramas, scrutinizing their narrative frameworks and emotional contours to discern their global resonance. It examines the cultural and emotional elements that strike a chord with international viewers, accentuating Türkiye's burgeoning role in the global entertainment domain. Through an amalgamation of sentiment analysis and narrative scrutiny, this research elucidates how these dramas foster emotional engagement and cultural connectedness, amplifying Türkiye's media influence worldwide. Drawing upon the narrative engagement and emotional investment theories posited by Vorderer and Klimmt \cite{VordererKlimmt2016}, this study ventures into the distinctive narrative rhythms that set Turkish storytelling apart from its global contemporaries, enriching the academic discourse on narrative psychology and media studies.

\section*{Diffusion Theory}
\label{sec:theory}
In this research, we delve into the dynamics influencing the popularity of cultural products, specifically focusing on Turkish TV series. The overarching question seeks to unravel the reasons behind the varying degrees of popularity among cultural products. This inquiry is further refined to Turkish TV series, examining firstly, the attributes that contribute to the popularity of certain series over others, termed the ``arc hypothesis," and secondly, the factors that determine the reception of Turkish series in different countries referred to as the "cultural affinity hypothesis."

Culture profoundly impacts our perspectives and expectations in life, structuring our daily experiences and shaping our preferences \cite{DiMaggio1997, Zerubavel1997}. As Bourdieu (1979) articulates, culture is a key determinant of our tastes \cite{bourdieu2018distinction}. Cultures can be differentiated based on their enduring orientations towards values and patterns of differentiation, as highlighted by Hofstede (1980) \cite{Hofstede1980}, Schwartz (1994) \cite{Schwartz1994}, and Welzel and Inglehart (2010) \cite{WelzelInglehart2010}, which in turn influence individual behaviors and tastes \cite{Benedict1934, Bourdieu1979}. From this understanding, we propose three conjectures: individual preferences reflect broader collective tastes, cultural products that align with these collective preferences are more likely to be well-received within that culture, and cultures with similar value systems tend to favor similar cultural products.

Drawing from the theoretical underpinnings of diffusion theories and the concept of cultural globalization, we posit the following hypotheses for our study:

\begin{itemize}
   \item[H1] Turkish TV series that exhibit Rise-Fall story arcs enjoy greater popularity, resonating with the narrative inclinations inherent in collective cultural tastes.
    \item[H2] The popularity of Turkish TV series across different countries is modulated by cultural affinity, suggesting that nations with social values that align with the thematic content of these series are more inclined to appreciate them. 
\end{itemize}

These hypotheses intertwine with the principles derived from diffusion theories on the spread of cultural products across structural, local, and individual layers, emphasizing the role of cultural practices and the innovative reinterpretation of global cultural ideas by local communities \cite{PopeMeyer2016, FourcadeGourinchasBabb2022, Mayer1997}. Moreover, the notion of cultural globalization accentuates the significance of cultural affinity in the acceptance and popularity of cultural products across diverse regions, bolstering our second hypothesis.

This research aims to enrich the understanding of the factors that shape the global dissemination and reception of cultural products by examining the case of Turkish TV series through these theoretical lenses.

\section*{Data and Method}
\label{sec:data}

A movie synopsis is a brief summary of the plot of a movie. It provides a concise description of the main events and characters in the story, often highlighting the central conflict and the development of the narrative. A well-crafted synopsis usually does not spoil key twists or the ending but aims to give enough information to attract interest while leaving the most important surprises and resolutions unrevealed. Synopses provide an overview of the plot, outlining basic elements of the story, including the setup, main characters, and central conflicts. It is a valuable marketing tool, attracting an audience by giving them a taste of the story. Industry professionals also use it to evaluate the content and potential of a film.

The data we are using contains the synopses of the Turkish series. 
We collected the titles of the dramas, and using these titles, we retrieved synopses from the Turkish-Drama.com database to gain an overview of each drama’s plot and genre. To capture information on release dates across different countries, we employed web-scraping techniques on IMDb, which provided a comprehensive record of when each drama was officially released or made available outside Türkiye.

\begin{itemize}
    \item \url{https://www.turkishdrama.com/} for the corpus of Turkish TV series
    \item Wikipedia and IMDB for the timestamp of international broadcast 
    \item Inglehart-Welzel Cultural Map of the world \url{https://www.worldvaluessurvey.org/WVSNewsShow.jsp?ID=467}
\end{itemize}

Our methodology integrates advanced sentiment analysis techniques with a novel corpus of Turkish TV series synopses sourced from turkishdrama.com. We supplement this with broadcast timestamps from IMDb and Wikipedia to construct a comprehensive dataset. The emotional valence of narrative segments is quantified through sentence-level sentiment analysis, enabling us to categorize the overarching emotional arcs of the narratives.

Sentiment analysis, traditionally used in various fields, is now being applied in literary studies \cite{mantyla2018evolution, boudad2018sentiment, drus2019sentiment, kumar2020systematic, ligthart2021systematic}. This work uses AI to match narratives with their ideal sentiment analysis model. This approach, combined with a focused textual analysis, offers fresh insights into the diverse narrative structures in literature.
For the analysis, we segment each plot into sentences and run sentiment polarity classification at the sentence level. We also segment each plot into three equal-length acts (beginning, middle, and end) following Field (1979) \cite{Field1979Screenplay}. These acts could vary in length and might occur in a different order, but we are assuming that is not the case for the sake of this exploratory project. We aggregate the sentiment polarity for each segment to determine the emotional arc of the plot. If the polarity is skewed toward positivity in the beginning but gradually tilts towards negativity in the middle as well as the end of the plot, we categorize this arc as a Rise-Fall arc. Its converse would be a Fall-Rise arc where the plot gradually moves from negativity to positivity.

\subsection*{Cultural Distance Measure}
For cultural distance, we relied on data provided by the World Values Survey (WVS). According to the WVS framework, there are two primary dimensions of cross-cultural variation: Traditional Values versus Secular-Rational Values and Survival Values versus Self-Expression Values. After extracting each country’s position on these dimensions, we standardized the scores and calculated the cultural distance from Türkiye. In our coding, a cultural distance of 0 indicates no divergence from Türkiye (i.e., Türkiye itself), whereas the most divergent country, Sweden, is coded as 1. All other countries fall between these two extremes based on their standardized cultural distance scores.

\subsection*{Geographical Distance}
To capture geographical distance, we measured the direct distance (in kilometers) from each country’s capital city to Ankara, the capital of Türkiye. By focusing on the distance between capital cities, we ensure consistency in the measurement across different countries.

\subsection*{Data Analysis: Poisson Regression}
Using the collected release data, we identified both:
The number of dramas diffused in each country.
The number of countries to which each drama was exported.
These counts allowed us to examine how widely Turkish dramas spread around the world. We employed a Poisson regression model to analyze how cultural and geographical distances from Türkiye influence the diffusion of Turkish dramas. Specifically, the dependent variable in the model is the count of dramas released in each country. In other words, for each country, we regress the number of introduced dramas in terms of both the cultural and geographical distance from Türkiye.

\subsection*{Sentiment Analysis and Emotional Arcs}
To explore the emotional trajectories in Turkish dramas, we drew on the emotional arc theory (Reagan et al. 2016), which posits six classic patterns of narrative emotion: Rise, Fall, Rise–Fall, Fall–Rise, Rise–Fall–Rise, and Fall–Rise–Fall.

We used a Large Language Model (LLM) to classify each drama according to the emotional arc that best describes its primary trajectory. Our findings revealed that the most common patterns among the analyzed dramas are Fall–Rise, Fall–Rise–Fall, and Rise–Fall–Rise. Because most dramas begin with a “fall” pattern, we combined all the narratives that begin with “rise” into a single category for comparison purposes.

\subsection*{Multinomial Logistic Regression}

Finally, we ran a multinomial logistic regression to investigate the relationship between the emotional arcs of these dramas and the extent of their international diffusion. In this model, the categorical outcome variable was the emotional arc classification, and the key explanatory variables were, again, the cultural and geographical distance to Türkiye. This analysis allows us to determine whether certain types of emotional arcs are more likely to be distributed in culturally or geographically distant markets.
Overall, this multi-faceted approach—integrating data collection, cultural and geographical distance measures, Poisson regression for diffusion, and multinomial logistic models for emotional arc patterns—provides a comprehensive understanding of how Turkish dramas spread internationally and how their narrative structures may align with different audiences around the globe.


\section*{Results}
\label{sec:results}
Our data show that the “successful” TV series we selected were shown on various TV channels. For example, Magnificent Century (Muhteşem Yüzyıl), a drama series with a focus on history and harem life, has been shown in 54 different countries since its release in Türkiye in 2011. A new production, Ömer, a drama series about a conservative family, released in 2023 on Turkish television, has already been released in at least five other countries since its release date. 

\subsection*{Story Arcs}
\begin{figure}
    \centering
\includegraphics[width=\linewidth]{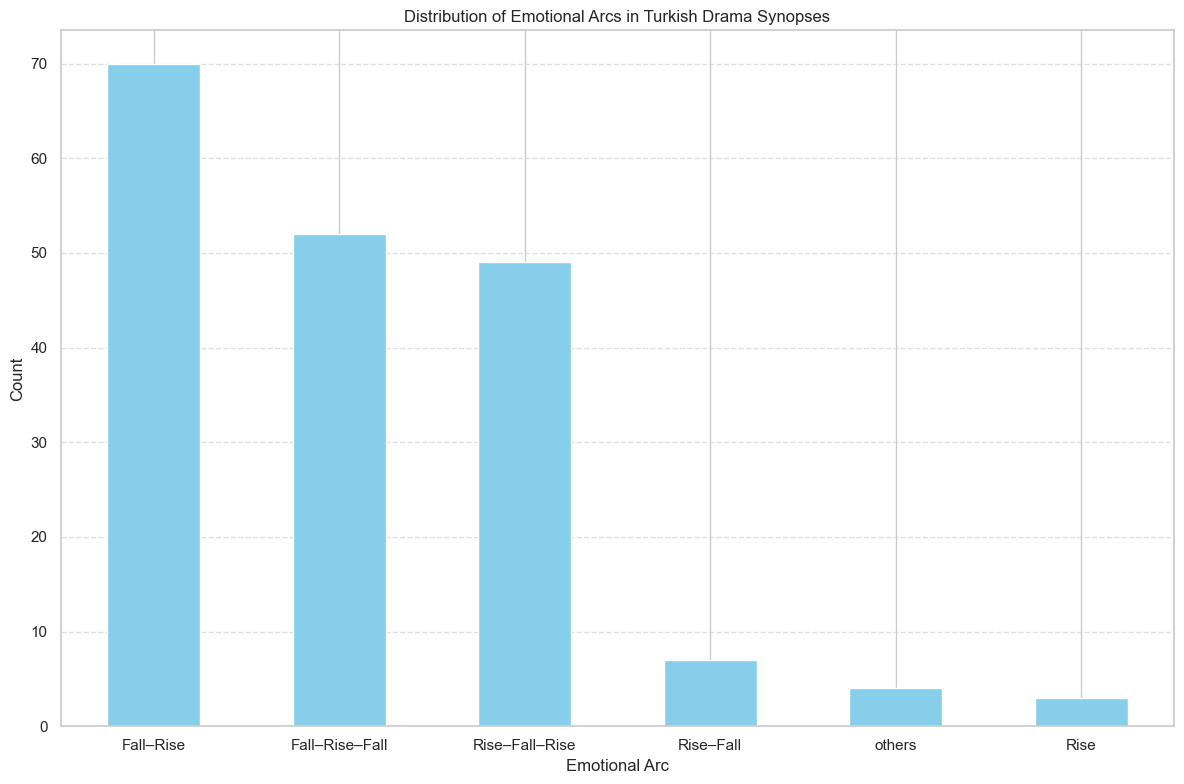}
    \caption{Distribution of Emotional Arcs in Turkish Drama Synopses}
    \label{fig:1}
\end{figure}

Fig.~\ref{fig:1} shows the analysis of the emotional trajectories in Turkish TV dramas revealed a predominant ``Fall-rise" narrative arc, characterized by an initial period of sadness that gradually gives way to more complex emotional states, with occasional peaks to maintain viewer interest. This pattern aligns with the ``Man in a Hole" arc commonly observed in Western narratives, suggesting a cultural specificity in Turkish storytelling.

Looking at the Turkish series, however, we can see that they cumulatively represent a time series with high values in early time, declining in time, having a bump at some middle point to give the audience a little hope that things may change and eventually fall again. So, our early investigation shows that in the Turkish series, which appeared in 2002, the trend went from a happy beginning to a somewhat sad ending with some minor fluctuations in between. 

\subsection*{Turkish series around the world}
\begin{figure}
    \centering
\includegraphics[width=\linewidth]{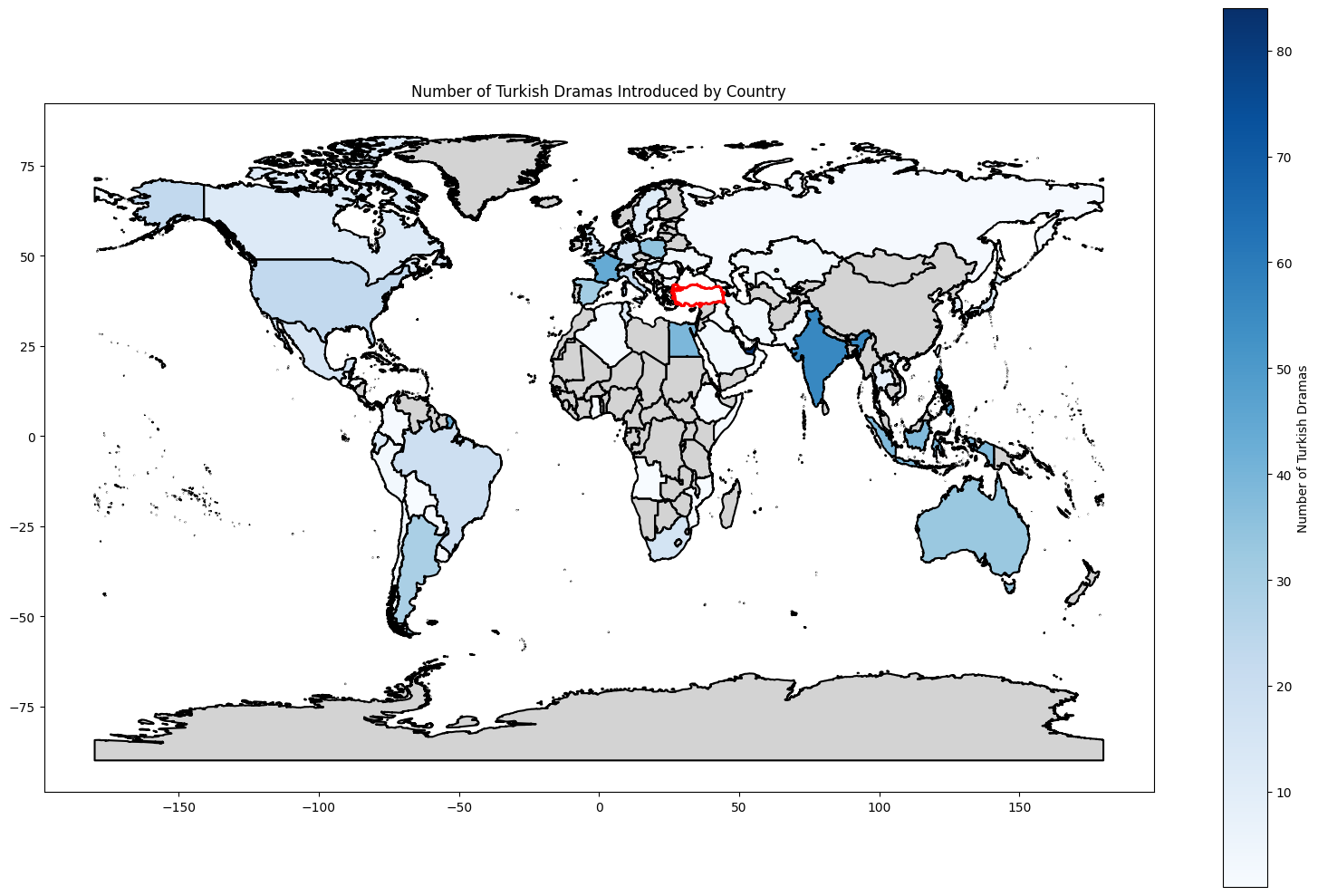}
    \caption{Number of Turkish Dramas Introduced by Country}
    \label{fig:2}
\end{figure}

\begin{figure}
    \centering
\includegraphics[width=\linewidth]{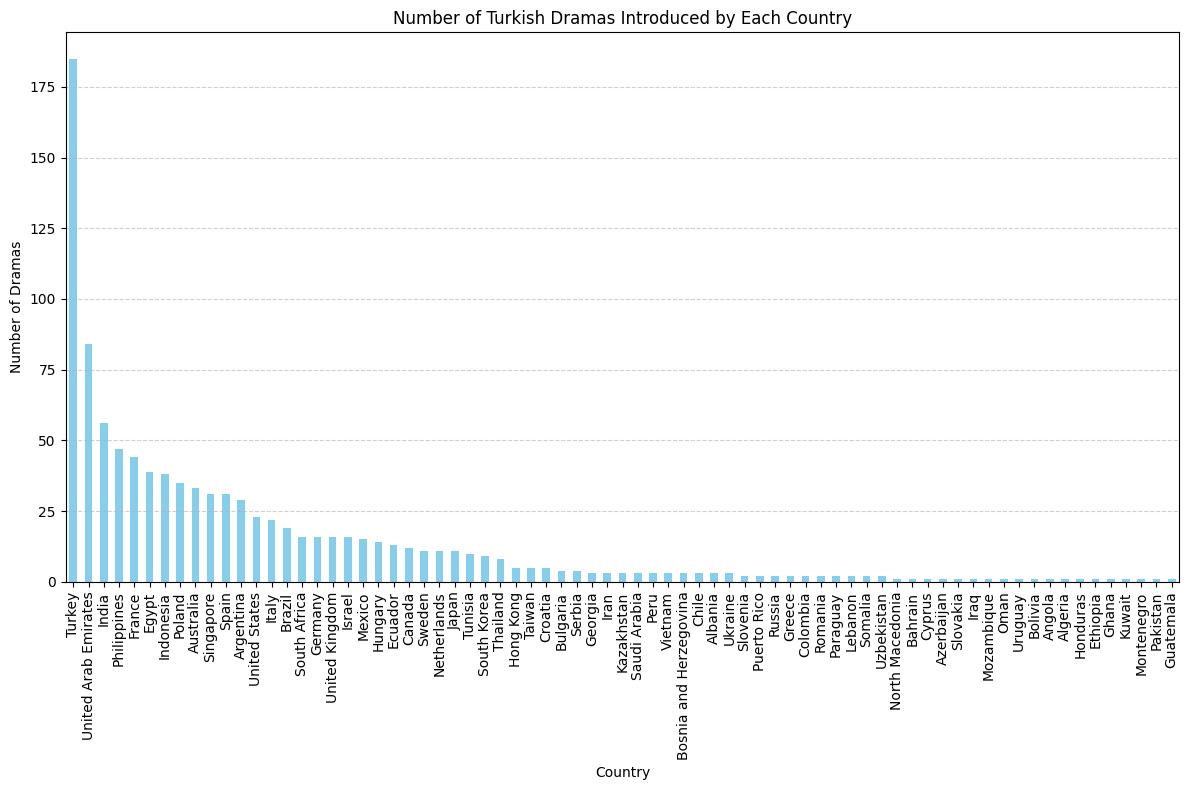}
    \caption{Number of Turkish Dramas Introduced by Each Country}
    \label{fig:3}
\end{figure}

\begin{figure}
    \centering
\includegraphics[width=\linewidth]{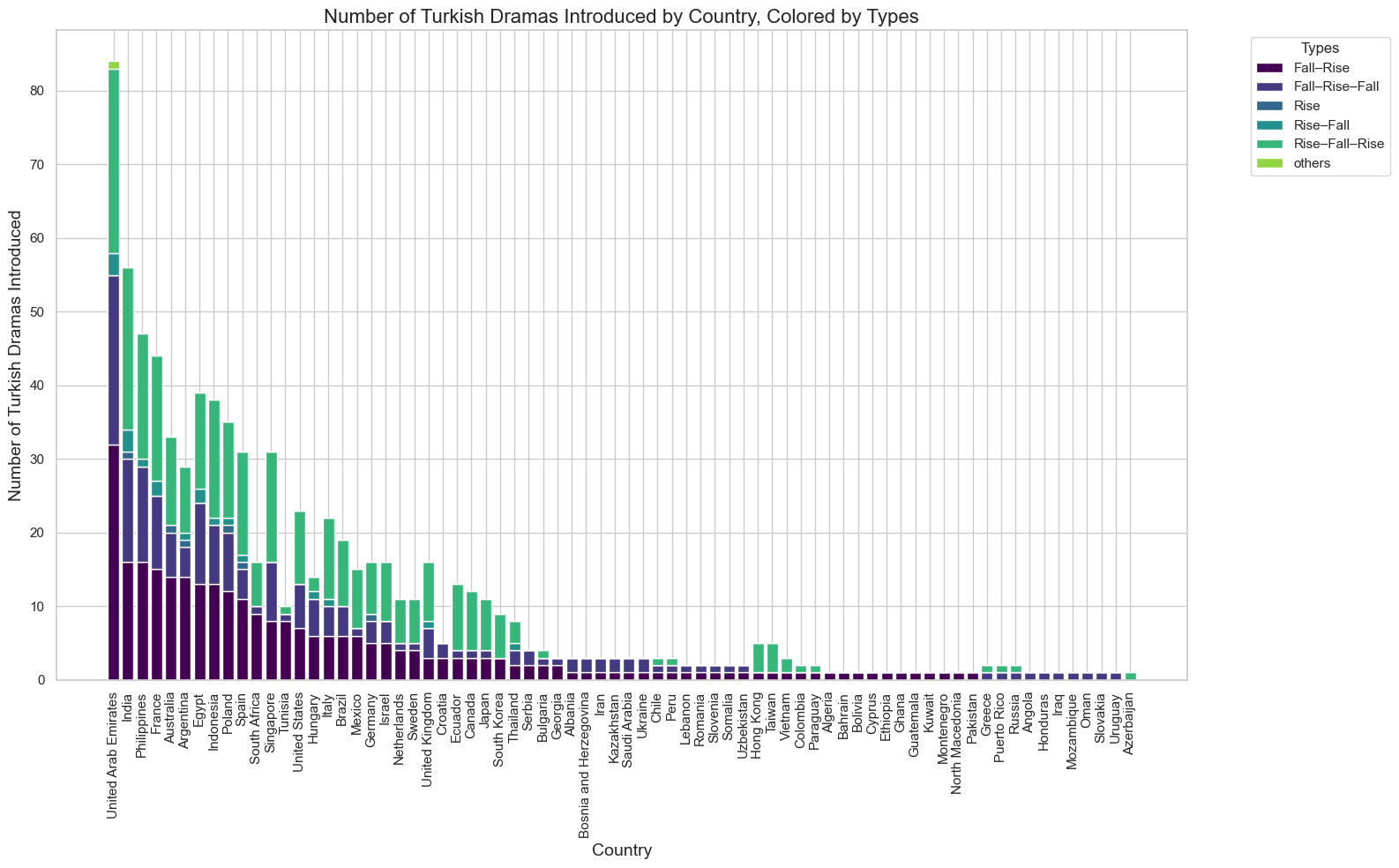}
    \caption{Number of Turkish Dramas Introduced by Country, Colored by Types}
    \label{fig:4}
\end{figure}

Our results, based on the 20 most famous Turkish TV series, show varying global distribution. Some series, which were also successful in Türkiye, had high demand levels from many countries worldwide. In contrast, others had a limited reach, mainly limited to Latin American countries and Balkan countries. Fig.~\ref{fig:2}, Fig.~\ref{fig:3}, and Fig.~\ref{fig:4}  show the number of Trukish series in each country. 

\begin{figure}
    \centering
    \includegraphics[width=\linewidth]{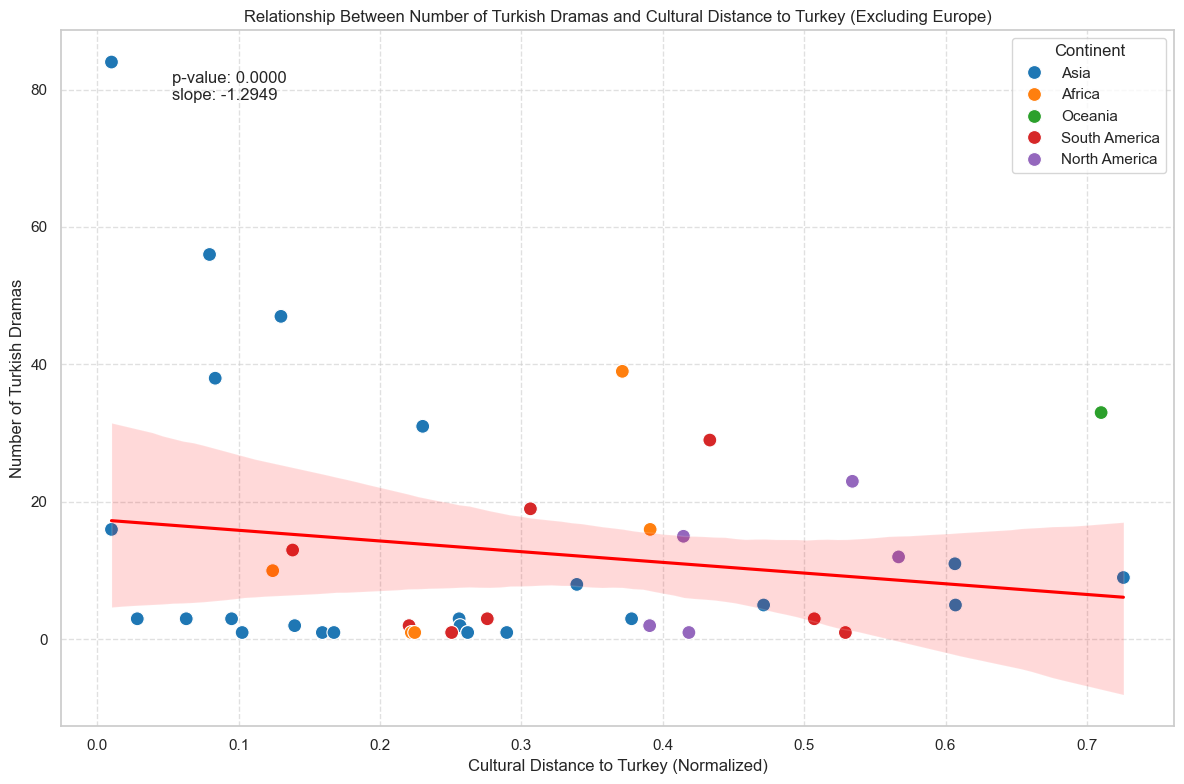}
    \caption{Relationship Between Number of Turkish Dramas and Cultural Distance to Türkiye (Excluding Europe)}
    \label{fig:5}
\end{figure}

\begin{figure}
    \centering
\includegraphics[width=\linewidth]{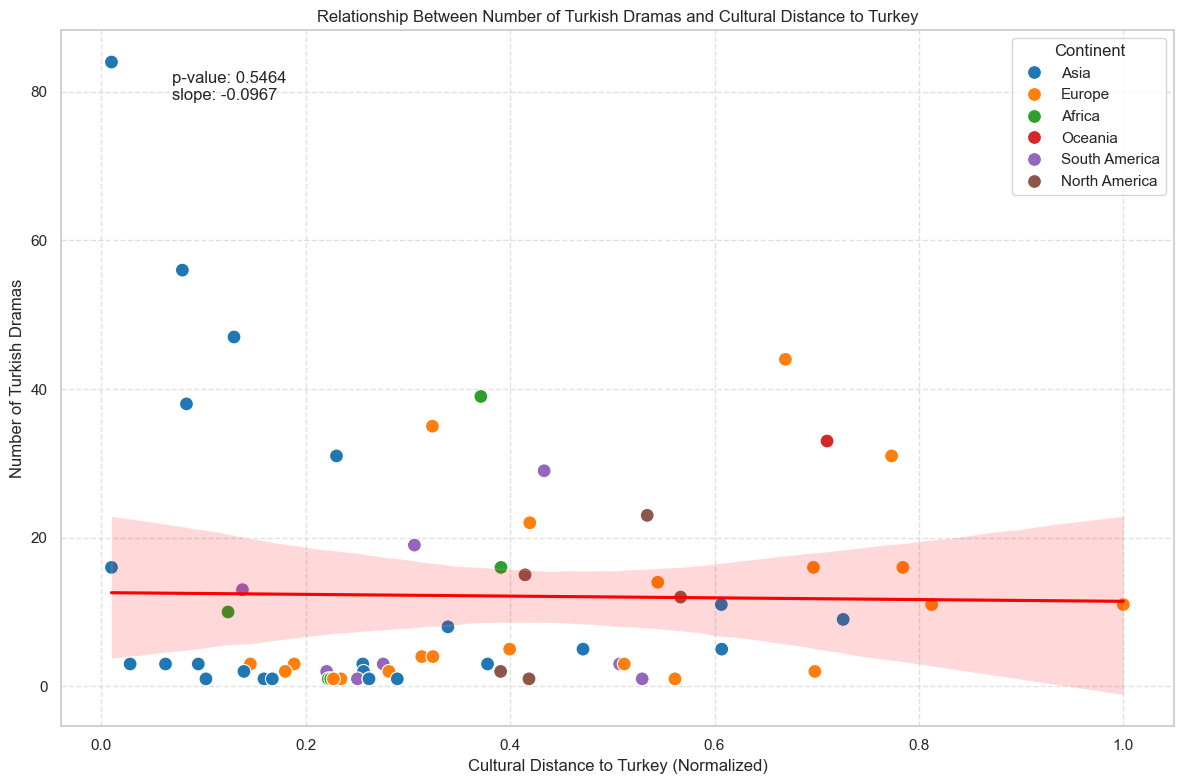}
    \caption{Relationship Between Number of Turkish Dramas and Cultural Distance to Türkiye}
    \label{fig:6}
\end{figure}

One reason these TV series might reach different countries earlier or later could be cultural similarity and shared preferences for similar cultural products. In Fig.~\ref{fig:5} and Fig.~\ref{fig:6}, we see that, indeed, Muhtesem Yuzyil was first released in countries that were spatially close, which are also the countries that have a shared history as well as similarities in traditional customs and consumer choices, such as Balkan countries and many Arab countries. Interestingly, we also see that several South American countries were also among the first consumers of his specific drama series around the world.

\section*{Conclusion \& Discussion}
\label{sec:discussion}
The paper investigates the success of Turkish TV dramas (Türk dizileri), focusing on their narrative structures and emotional arcs. It discusses Türkiye's rise as a global entertainment influencer, with these series portraying Turkish culture authentically and resonating across various regions. The study uses sentiment analysis to categorize the emotional journey of stories and examines the popularity of these dramas in different countries, considering cultural similarities and preferences. The research aims to understand why some Turkish series are more popular than others and their global appeal.

There are many interesting questions to look at here:

\begin{itemize}
    \item Looking at the times series of each story and the cumulative one, we can see that the tail of the series either curves up or down. This means that at the end of each series, there is a significant change in the story; things change and give the audience more momentum. Now, the question is to quantify the tail and investigate to what extent it is a thing! How common is this pattern? This pattern also exists in Peter Dodds's figures (90 -- 100 \%). Does the tail tell us more? Like a clue when to end a series or a movie?

    \item The Data we use has a limited amount of granularity. Our educated guess is that we have the script for every episode; it should have a rough surface, many small fluctuations with large ones mainly at the end of each episode, and some high peaks only in some episodes. 

    \item Conservation of emotion (area under the curves)

\end{itemize}

\section*{Acknowledgments}
AKR acknowledges support from the Independent Research Fund Denmark (EliteForsk grant to Sune Lehmann), the Carlsberg Foundation (the Hope Project) and the Villum Foundation (NationScale Social networks). Some of the simulations presented above were performed using computer resources within the Aalto University School of Science ``Science-IT'' project. This study's simulations and numerical computations are publicly available at \cite{simulations}.
\nolinenumbers

%
%
\bibliographystyle{plos2015}  
\bibliography{citations}

\begin{thebibliography}{10}

\bibitem{berg2023turkish}
Berg M.
\newblock Turkish Drama Serials: The Importance and Influence of a Globally Popular Television Phenomenon.
\newblock University of Exeter Press; 2023.

\bibitem{vonnegut1999palm}
Vonnegut K.
\newblock Palm Sunday: an autobiographical collage.
\newblock Dial Press Trade Paperback; 1999.

\bibitem{reagan2016emotional}
Reagan AJ, Mitchell L, Kiley D, Danforth CM, Dodds PS.
\newblock The emotional arcs of stories are dominated by six basic shapes.
\newblock EPJ Data Science. 2016;5(1):1--12.

\bibitem{del2021improving}
Del~Vecchio M, Kharlamov A, Parry G, Pogrebna G.
\newblock Improving productivity in Hollywood with data science: Using emotional arcs of movies to drive product and service innovation in entertainment industries.
\newblock Journal of the Operational Research Society. 2021;72(5):1110--1137.

\bibitem{VordererKlimmt2016}
Vorderer P, Klimmt C.
\newblock Narratives, the Embodiment of Stories, and Their Impact on the Human Mind.
\newblock In: Reinecke L, Oliver MB, editors. The Routledge Handbook of Media Use and Well-Being. Routledge; 2016. p. 7--18.

\bibitem{DiMaggio1997}
DiMaggio P.
\newblock Culture and Cognition.
\newblock Annual Review of Sociology. 1997;23:263--287.

\bibitem{Zerubavel1997}
Zerubavel E.
\newblock Social Mindscapes: An Invitation to Cognitive Sociology.
\newblock Harvard University Press; 1997.

\bibitem{bourdieu2018distinction}
Bourdieu P.
\newblock Distinction a social critique of the judgement of taste.
\newblock In: Inequality. Routledge; 2018. p. 287--318.

\bibitem{Hofstede1980}
Hofstede G.
\newblock Culture's Consequences: International Differences in Work-Related Values.
\newblock Beverly Hills, CA: Sage. 1980;.

\bibitem{Schwartz1994}
Schwartz SH.
\newblock Beyond Individualism/Collectivism: New Cultural Dimensions of Values.
\newblock Sage Publications, Inc. 1994;.

\bibitem{WelzelInglehart2010}
Welzel C, Inglehart R.
\newblock Agency, Values, and Well-Being: A Human Development Model.
\newblock Social Indicators Research. 2010;97(1):43--63.

\bibitem{Benedict1934}
Benedict R.
\newblock Patterns of Culture.
\newblock Houghton Mifflin; 1934.

\bibitem{Bourdieu1979}
Bourdieu P.
\newblock Distinction: A Social Critique of the Judgement of Taste.
\newblock Harvard University Press; 1979.

\bibitem{PopeMeyer2016}
Pope S, Meyer JW.
\newblock Universal Legitimacy: Theories and Concepts.
\newblock Journal of Global Sociology. 2016;.

\bibitem{FourcadeGourinchasBabb2022}
Fourcade-Gourinchas M, Babb SL.
\newblock Global Patterns and the Spread of Ideas.
\newblock International Review of Sociology. 2022;.

\bibitem{Mayer1997}
Mayer KU.
\newblock The Organized Hierarchy of Power and Interests.
\newblock Academic Press; 1997.

\bibitem{mantyla2018evolution}
M{\"a}ntyl{\"a} MV, Graziotin D, Kuutila M.
\newblock The evolution of sentiment analysis—A review of research topics, venues, and top cited papers.
\newblock Computer Science Review. 2018;27:16--32.

\bibitem{boudad2018sentiment}
Boudad N, Faizi R, Thami ROH, Chiheb R.
\newblock Sentiment analysis in Arabic: A review of the literature.
\newblock Ain Shams Engineering Journal. 2018;9(4):2479--2490.

\bibitem{drus2019sentiment}
Drus Z, Khalid H.
\newblock Sentiment analysis in social media and its application: Systematic literature review.
\newblock Procedia Computer Science. 2019;161:707--714.

\bibitem{kumar2020systematic}
Kumar A, Jaiswal A.
\newblock Systematic literature review of sentiment analysis on Twitter using soft computing techniques.
\newblock Concurrency and Computation: Practice and Experience. 2020;32(1):e5107.

\bibitem{ligthart2021systematic}
Ligthart A, Catal C, Tekinerdogan B.
\newblock Systematic reviews in sentiment analysis: a tertiary study.
\newblock Artificial Intelligence Review. 2021; p. 1--57.

\bibitem{Field1979Screenplay}
Field S.
\newblock Screenplay: The Foundations of Screenwriting.
\newblock Dell Publishing; 1979.

\bibitem{simulations}
Github Repository;.
\newblock \url{https://github.com/pp33pant/turkish_arcs}.

\end{thebibliography}

\end{document}